\documentclass{article}
\usepackage{ijcai21}

\usepackage{times}
\usepackage{soul}
\usepackage{url}
\usepackage[hidelinks]{hyperref}
\usepackage[utf8]{inputenc}
\usepackage[small]{caption}
\usepackage{graphicx}
\usepackage{amsmath}
\usepackage{amsthm}
\usepackage{booktabs}
\usepackage[linesnumbered,ruled,vlined]{algorithm2e}
\usepackage{mathrsfs}
\usepackage{algorithmic}
\usepackage{amsmath,amssymb} 
\usepackage{subfig}
\usepackage{epsfig}
\usepackage{multirow}
\usepackage{here}
\usepackage{amsthm}
\usepackage{tikz}

\newcommand{\graph}{\ensuremath{\mathcal{G}}}
\newcommand{\steiner}{\ensuremath{\mathcal{S}}}
\newcommand{\vertices}{\ensuremath{\mathcal{V}}}
\newcommand{\edges}{\ensuremath{\mathcal{E}}}
\newcommand{\terminals}{\ensuremath{\mathcal{T}}}

\newcommand{\frontiers}{\ensuremath{\mathcal{F}}}
\newcommand{\threshold}{\theta}

\newcommand{\bdd}{\ensuremath{\mathcal{D}}}
\newcommand{\nodes}{\ensuremath{\mathcal{N}}}
\newcommand{\arcs}{\ensuremath{\mathcal{A}}}
\newcommand{\mst}{\ensuremath{\mathcal{M}}}

\theoremstyle{definition}
\newtheorem{definition}{Definition}
\newtheorem{theorem}{Theorem}

\newtheorem{lemma}{Lemma}

\makeatletter
\def\Hline{
  \noalign{\ifnum0=`}\fi\hrule \@height 3.\arrayrulewidth \futurelet
  \reserved@a\@xhline}
\makeatother

\title{Cost-constrained Minimal Steiner Tree Enumeration by Binary Decision Diagram}

\author{
     Yuya Sasaki
    \affiliations
     Osaka university
    \emails
     sasaki@ist.osaka-u.ac.jp
}

\begin{document}

\maketitle

\begin{abstract}
The Steiner tree enumeration problem is a well known problem that asks for enumerating Steiner trees.
Numerous theoretical works proposed algorithms for the problem and analyzed their complexity, but there are no practical algorithms and experimental studies.
In this paper, we study the Steiner tree enumeration problem practically. 
We define a problem {\it cost-constrained minimal Steiner tree enumeration problem}, which enumerates minimal Steiner trees with costs not larger than a given threshold. 
To solve the problem, we propose a binary decision diagram (BDD)-based algorithm.
The BDD-based algorithm constructs a BDD that compactly represents the set of minimal Steiner trees and then traverses the BDD for enumeration. 
We develop a novel {\it frontier-based algorithm} to efficiently construct BDDs.
Our BDD traverse algorithm prunes Steiner trees with costs larger than the threshold.
We also extend our algorithm by preprocessing the given graph and controlling the number of generated Steiner trees in order to reduce the memory and computation costs.
The extension makes our algorithm scalable by generating a subset of the minimal Steiner trees.
We validate that our algorithm can enumerate Steiner trees in real-world graphs more efficiently than existing methods.

\end{abstract}

\section{Introduction}
The {\it Steiner tree problem} is one of the most classic and fundamental problems for graph analysis~\cite{dreyfus1971steiner}.
The problem asks for the minimum-cost {\it Steiner tree} which is a subtree that connects an arbitrary subset of vertices (called {\it terminals}) in a given graph.
The Steiner tree problem emerges in numerous applications such as a design of VLSI, network routing, road network construction, and wire length estimation \cite{hwang1992steiner} though it is known as NP-hard problem \cite{garey1979computers}.

The minimum-cost Steiner tree may not be the best solution in real-world applications because the applications implicitly may have constraints that Steiner trees should have.
It is possible to add such constraints to objective functions but it is often difficult to formalize them.
For example, let us consider a road network construction. Road networks are modeled by graphs whose vertices represent intersections/facilities and edges represent road segments.
For robustness to disasters such as earthquakes and flooding, users often want to know Steiner trees such that roads are ``not very close'' to each other \cite{tsompanakis2008structural}.
However, it is difficult to set an appropriate distance threshold and/or formalize distances between edges. 


The {\it Steiner tree enumeration problem} is a problem that asks for enumerating Steiner trees in a given graph \cite{dourado2014algorithmic,kimelfeld2008efficiently}.
We can find a suitable Steiner tree from enumerated Steiner trees. 
The problem is theoretically studied well in terms of time and memory complexity. 
However, to the best of our knowledge, there do not exist practical algorithms and experimental studies.
To practically solve the Steiner tree enumeration problem, we can use (approximate or exact) algorithms for the Steiner tree problem. 
In more concretely, a straightforward method for enumerating Steiner trees is that iteratively (1) runs algorithms to generate  Steiner trees and (2) modify a given graph to avoid obtaining the same Steiner tree, until enumerating all Steiner trees.
The straightforward method has two issues; (1) how to modify the graph for enumerating Steiner trees and (2) when the method can be terminated. 
We need a practical problem definition and efficient methods.
In this paper, we initiate a new problem for enumerating Steiner trees, which we call {\it cost-constrained minimal Steiner tree enumeration problem}.
The problem asks for enumerating minimal Steiner trees with costs not larger than a given threshold.
We theoretically show the problem is $\#$P-hard problem.
For efficient enumeration, we propose a novel algorithm that enumerates Steiner trees by using binary decision diagram (BDD for short).
BDD is a well-known data structure for effective enumeration of discrete data including graphs.
Our BDD-based algorithm has two steps; constructing a BDD and traversing the constructed BDD.
For the first step, we develop a novel {\it frontier-based algorithm} to efficiently construct BDDs that compactly represent the set of minimal Steiner trees.  
The frontier-based algorithm effectively reduces the size of BDDs by using functions tailored for the Steiner tree enumeration problem. 
After constructing a BDD, we traverse the BDD to obtain the set of Steiner trees while pruning the Steinter trees with larger costs than the threshold.

It is generally difficult to construct BDDs for large-scale graphs due to time and space complexity. Recent works for enumerating subgraphs showed that computational costs and memory usage become significantly large even for graphs with few hundred edges \cite{maehara2017exact,sasaki2019efficient}.
Thus, we restrict the number of generated Steiner trees by applying preprocessing techniques and extending the traverse algorithm.
The preprocessing technique reduces the size of graph by using existing algorithms for the Steiner tree problem. 
The traverse algorithm enumerates the Steiner trees with the $k$ smallest costs among Steiner trees on preprocessed graphs, where $k$ is a user-specified parameter.
The extension makes our algorithm scalable by generating a subset of the minimal Steiner trees.

We conduct computational experiments to evaluate our BDD-based algorithm by using nine real graphs.
We show that our algorithm efficiently enumerates Steiner trees compared with graph traversal-based and solver-based methods.
Furthermore, the enumerated Steiner trees have costs close to the optimal cost.




\section{Preliminaries}\label{sec:preliminaries}
We describe the notations and definitions used in this paper.

\subsection{Definitions and Notations}
Let $\graph =(\vertices,\edges)$ be an undirected, weighted, and connected graph, where 
$\vertices$ and $\edges = \vertices \times \vertices$ are sets of vertices and edges, respectively.
Each edge $e$ has a non-negative weights $c(e)$. 
Each vertex has an unique integer $v$ as its identifier.

\begin{definition}[Steiner Tree]
A subset of edges $\steiner \subseteq \edges$ spanning a given set of terminals $\terminals \subseteq \vertices$ without cycles is called a Steiner tree, and the cost of $\steiner$ is $c(\steiner) = \sum_{e \in \steiner} c(e)$.
\end{definition}

Given Steiner tree $\steiner$, a subgraph $\steiner \cup e$ could be a Steiner tree. However, the subgraph is meaningless if we know $\steiner$ already.
We define {\it minimal Steiner tree} as follows:

\begin{definition}[Minimal Steiner tree]
Given Steiner tree $\steiner$, $\steiner$ is a minimal Steiner tree if and only if there are no $e$ in $\steiner$ such that $\steiner \backslash e$ is a Steiner tree.  
\end{definition}
\noindent We here note that Steiner trees whose all leaves are terminals are minimal.

\begin{definition}[Steiner tree problem]
Giving a graph and a set of terminals, the Steiner tree problem asks for the Steiner tree with the minimum cost.
\end{definition}

 \subsection{Binary Decision Diagram}

To maintain exponentially numerous subgraphs, we can use the binary decision diagram (BDD), which is a data structure to represent a Boolean function compactly.

A BDD  $\bdd =(\nodes, \arcs)$ is a directed acyclic graph with sets of nodes $\nodes$ and arcs $\arcs$\footnote{To avoid confusion, we use the terms ``vertex'' and ``edge'' to refer to a vertex and an edge in a graph, respectively, and ``node'' and ``arc'' to refer to a vertex and an edge in a BDD, respectively.}.
It has two nodes that have no outgoing arcs, called {\it sink nodes}.
The sink nodes are of two types, called {\it 1-sink} and {\it 0-sink}. 
Additionally, it has a single node that has no incoming arcs, called the {\it root node}.
Each node (except for sink nodes) has two outgoing arcs, called the {\it 0-arc} and {\it 1-arc}.
Each node is assigned a variable, and 0-arc and 1-arc indicate that the variable of the node is false and true, respectively.
We call the set of descendent nodes through $x$-arc from node $n$ as $x$-descendants of $n$.

Paths from the root node to the 1-sink and 0-sink represent variable assignments that a given Boolean function is true and false, respectively.
In the contexts of graph enumeration, each variable represents an edge. A path from the root node to the 1-sink represents a constrained subgraph that is represented by a Boolean function.

The size of the BDD is defined by the number of nodes in the BDD \cite{hardy2007k}. 
Generally, it exponentially increases as edges in the graphs increase.
As the sizes of graphs increase, both of the computation cost and the memory usage exponentially increase.
Thus, it is difficult to construct BDD on large-scale graphs. 

\section{Proposal}\label{sec:proposal}
We aim at enumerating minimal Steiner trees instead of finding the optimal Steiner tree.
We define a new problem that we solve in this paper as follows:

\smallskip
\noindent
{\bf Problem definition} ({\it Cost-constrained Minimal Steiner Tree Enumeration Problem})
Given a weighted undirected graph $\graph=(\vertices,\edges)$, a subset of its vertices $\terminals \subseteq \vertices$, and a positive threshold $\threshold$,
the cost-constrained minimal Steiner enumeration problem asks for enumerating minimal Steiner trees whose costs are not larger than $\threshold$.
\smallskip

The Steiner tree problem is known as NP-hard problem.
While, the minimal Steiner tree enumeration problem is $\#P$-hard problem as follows:
\begin{theorem}
The minimal Steiner tree enumeration problem is $\# P$-hard problem.
\end{theorem}
{\it Proof sketch}: We reduce the enumeration of $s$-$t$ simple pats~\cite{valiant1979complexity} which is $\# P$-hard problem to the minimal Steiner tree enumeration problem.
The minimal Steiner tree enumeration problem can formulate the enumeration of $s$-$t$ simple paths.
Given two vertices as terminals, the minimal Steiner tree enumeration problem is equivalent to the enumeration of $s$-$t$ simple paths.
Therefore, the minimal Steiner tree enumeration problem is $\#$P-hard problem. \hfill{} $\square$


In the following, we show an overall idea of our algorithm.
Then, we present our techniques; BDD construction, BDD traversal, and preprocessing methods.

\subsection{Overall Idea}
For thoroughly enumerating minimal Steiner trees, we need to search for the minimum-cost Steiner trees on all possible graphs, which are subgraphs of a given original graph. 
This is equivalent to find all variable assignments for which the Boolean function representing minimal Steiner trees is true.
Since the BDD can efficiently and compactly represent the Boolean function, it is suitable for enumerating minimal Steiner trees.
There are no studies on BDDs for enumerating minimal Steiner trees, and thus we design a novel algorithm for constructing the BDD.

After constructing BDDs, we traverse the constructed BDDs to enumerate the minimal Steiner trees. Since BDDs may include minimal Steiner trees with costs larger than the threshold, our traversal algorithm prunes such Steiner trees on demand.






\subsection{BDD Construction Method}

Our BDD construction method is a type of {\em frontier-based search}, which is a general construction algorithm for enumerating all constrained subgraphs \cite{kawahara2017frontier,maehara2017exact}.
The frontier-based search is applied to many types of subgraphs such as trees, connected graphs, and paths, but not for minimal Steiner trees. The pseudo-code of BDD construction method is in appendix.

We extend the frontier-based search for enumerating minimal Steiner trees.
The non-trivial challenge is how to design functions and frontier information of the frontier-based algorithm to reduce the size of BDD with keeping accurate Steiner tree enumeration.
We first describe the framework of the frontier-based search and then each function.

{\bf Frontier-based search}: 
The frontier-based search generates nodes of BDDs one by one.
We fix ordering of edges ($e_1,\ldots,e_{|\edges|}$) before constructing BDD and process the edges following the order.
The processed and unprocessed edges at the end of $i$-th step are denoted by $E^{\leq i} := \{e_1,\ldots,e_i \}$ and $E^{> i} := \{e_{i+1},\ldots,e_{|\edges|} \}$, respectively.
The set of vertices that have both processed and unprocessed edges is called the {\em frontier} (at the $i$-th step) and denoted by $\frontiers_i$.

We consider all minimal Steiner trees $\mst$.
The set of nodes at $i$-th step $\nodes_i$ represents all subsets of $E^{\leq i}$ that can possibly belong to $\mst$.
$\mst(n)$ denotes the set of edges that can possibly belong to $\mst$ represented by node $n$.
A path from the root node to $n$ represents a subgraph in which edges are present if the path descends the 1-arc of nodes associated with edges.

At the $i$-th step, the algorithm generates $\nodes_i$ from $\nodes_{i-1}$.
For each node $n \in \nodes_{i-1}$, the frontier-based search generates two children for which $e_i$ is excluded or included in the sets in $\mst$ by {\sf generateNode} function.
Generated nodes in $\nodes_i$ can be merged if the descendants of the nodes go to the same sinks for any patterns of $E^{> i}$ by {\sf mergeNode} function.
In addition, the frontier-based algorithm has two functions {\sf isZeroSink} and  {\sf isOneSink} to determine $x$-descendants of $n$ do not satisfy and do satisfy minimal Steiner trees, respectively, which output either true or false according to $x$ and information of $n$.

In the following, we present an extension of the frontier-based algorithm for minimal Steiner trees.
We first explain information held at each node, and then explain four functions; {\sf generateNode}, {\sf mergeNode}, {\sf isZeroSink}, and {\sf isOneSink}.

{\bf Frontier information}: 
Our BDD captures the characteristics of minimal Steiner trees for efficient enumeation.
Minimal Steiner trees are subgraphs that (1) connect all terminals (i.e. connectivity), (2) do not have cycles (i.e., tree), and (3) do not have edges that the subgraphs without the edges satisfy both (1) and (2) (i.e., minimality). Additionally, we aim to enumerate Steiner trees with costs not larger than $\threshold$.
Based on the above constraints, we design the information at node $n \in \nodes_i$ as follows:

\begin{itemize}
\item Identifier $id_n(f)$ for all $f \in \frontiers_i$:
If frontiers $f$ and $f'$ are connected by processed edges, $id_n(f)$ and $id_n(f')$ share the same identifier.
\item Unprocessed edge $u_n(f)$ for all $f \in \frontiers_i$: The sum of the numbers of unprocessed
edges connected to the frontiers such that $\{f \in \frontiers_i |id_n(f) =
id_n(f')\}$.
\item Terminal $t_n(f)$ for all $f \in \frontiers_i$: The number of terminals that are connected to $f$ by processed edges.
\item Degree $d_n(f)$ for all $f \in \frontiers_i$: The number of present edges that are connected to $f$.
\item Cost $c_n$: The cost of the subgraph corresponding to $n$.
\end{itemize}

We here note that if $id_n(f) = id_n(f')$, we hold  that $u_n(f) = u_n(f')$ and $t_n(f) = t_n(f')$ because $f$ and $f'$ are connected by processed edges. 
The information fulfills to check whether subgraphs represented by $n$ satisfy the conditions of minimal Steiner trees or never satisfy the conditions even after processing $E^{> i}$.
We extend {\sf generateNode}, {\sf mergeNode}, {\sf isZeroSink}, and {\sf isOneSink} functions based on the information.
In the following, let us assume that we process edge $e_i$ that connects to $v$ and $v'$.

{\bf generateNode function}:
The {\sf generateNode} function is to generate new nodes at $i$-th step and compute the information of the new nodes.
The information of new node $n'$ is first copied from its parent nodes $n$ and then updates according to that $e_i$ is either included or excluded.

We describe the detailed procedures in each case that $e_i$ is included and excluded.
As a common procedure, we add (resp. delete) the information for new (resp. old) frontier if a vertex is added to (resp. deleted from) frontiers. The information on the new frontier is derived from the graph. In the following, we assume that both $v$ and $v'$ are frontiers. If they are not frontiers, we can just ignore the updating of information related to them.
When we exclude $e_i$, we decrease $u_n(v)$ and $u_n(v')$ by one.
When we include $e_i$, we update the information as follows: $id_{n'}(v) = min(v,v')$, $u_{n'}(v) = u_{n}(v)+u_{n}(v')-2$, $d_{n'}(v) = d_{n}(v)+1$, and $c_{n'} = c_n + c(e_i)$. Similarly, we update the information related to $v'$.

{\bf mergeNode function}:
The {\sf mergeNode} function computes whether two nodes can be merged or not. Descendants of merged nodes go to the same 0-sink and 1-sink for any patterns of $E^{>i}$.
If two nodes $n$ and $n'$ in $\nodes_i$ have the following information, we merge $n$ and $n'$.

\begin{lemma}
Two nodes $n$ and $n'$ in $\nodes_i$ are merged if they have the following information: for all $f \in \frontiers_i$, $id_n(f) = id_{n'}(f)$, ($t_n(f)>0$ and $t_{n'}(f)>0$) or ($t_n(f)=0$ and $t_{n'}(f)=0$), and $d_n(f) = d_{n'}(f)$.
\end{lemma}
{\it Proof}:
We prove the theorem by each condition.
First, if $id_n(f) = id_{n'}(f)$, sets of connected frontiers on $n$ and $n'$ are same.
Descendants of $n$ and $n'$ are similarly connected all frontiers after $i$-th step.
Second, if ($t_n(f)>0$ and $t_{n'}(f)>0$) or ($t_n(f)=0$ and $t_{n'}(f)=0$), frontiers that must be connected are the same in order to connect all terminals. 
Third, if $d_n(f) = d_{n'}(f)$, both descendants of $n$ and $n'$ become either minimal or not minimal after $i$-th step.
If $n$ and $n'$ satisfy three conditions, these descendants go to the same sinks for any patterns of $E^{>i}$. Therefore, we can merge two nodes without missing any minimal Steiner trees. 
\hfill{} $\square$


If two nodes are merged, we use frontier information at either node $n$ or $n'$. 
While, $c_n$ is updated to the smaller ones of $c_n$ and $c_n'$ to avoid false-negative of $\mst$. That is, if $c_n$ is updated to the larger one, some Steiner trees may be discarded due to overestimating the costs of Steiner trees. 
On the other hand, our update may include false-positive due to underestimating costs, but we can remove such Steiner trees during traversing BDDs.

{\bf isZeroSink and isOneSink functions}: 
If subgraphs represented by nodes satisfy the conditions of minimal Steiner tree, the nodes should point at 1-sink. Similarly, if subgraphs represented by nodes do not satisfy the conditions of minimal Steiner tree ever, they should point at 0-sink.
We design {\sf isOneSink} and {\sf isZeroSink} functions to determine whether child nodes of $n$ are 1- and 0-sinks, or not. We explain ${\sf isOneSink}(n,e_i,x)$ and ${\sf isZeroSink}(n,e_i,x)$ such that $x$-arc from node $n$ that is assigned $e_i$ point at whether 1-sink and 0-sink or not, respectively.

First, the {\sf isOneSink} function is designed as follows:
\vspace{-1mm}
\[
  {\sf isOneSink}(n,e_i,1) = \left\{ \begin{array}{ll}
    true & \exists f\in \frontiers_i~~ t_n(f) = |\terminals| \\
     false & otherwise
  \end{array} \right.
\]

${\sf isOneSink}(n,e_i,0)$ is always false because if $e_i$ is excluded, all terminals are not connected on $n$ at $i$-th step.
Intuitively, {\sf isOneSink} function checks whether the all terminals are connected

There are several conditions of the {\sf isZeroSink} function, so we list the case that {\sf isZeroSink} is true;

${\sf isZeroSink}(n,e_i,1)$ becomes true if one of the followings satisfies;
\begin{itemize}
\item $id_n(v) = id_n(v')$,
\item $v \notin \frontiers_i$, $d_n(v)=0$ and $v \notin \terminals$ (similarly, for $v'$), and
\item $c_n + c(e_i) > \theta$.
\end{itemize}

${\sf isZeroSink}(n,e_i,0)$ becomes true if one of the followings satisfies;
\begin{itemize}
\item $v \not\in \frontiers_i \cup \frontiers_{i-1}$ and $v \in \terminals$ (similarly, for $v'$),
\item $t_n(v)>0$ and $u_n(v) = 1$ (similarly, for $v'$), and
\item $v \notin \frontiers_i$, $d_n(v)=1$ and $v \notin \terminals$ (similarly, for $v'$).
\end{itemize}

Intuitively, {\sf isZeroSink} function checks whether (1) terminals are disconnected, (2) subgraphs have cycle, and (3) subgraphs are not minimal.
{\sf isOneSink} and {\sf isZeroSink} functions help to reduce the number of nodes in the BDD.
We here note that both functions are true at the same time.

\begin{lemma}
${\sf isOneSink}(n,e_i,x)$ and ${\sf isZeroSink}(n,e_i,x)$ functions are true if $x$-arc from node $n$ that is assigned $e_i$ satisfies and does not satisfy conditions of minimal Steiner trees, respectively.
\end{lemma}
{\it Proof}:  
Remind that we obtain subgraphs by traversing from the root node to $n$. If $x$ is 1, we add edge $e_i$ to the subgraphs and if $x$ is 0, we delete $e_i$ from the subgraphs. 
If the subgraphs are minimal Steiner trees, they satisfy conditions that (1) connect all terminals (i.e., connectivity), (2) do not have cycles (i.e., tree), and (3) do not have edges that if the edges  are deleted,  subgraphs  satisfy  both  (1)  and  (2) (i.e., minimality).

We consider the four cases; (a) ${\sf isOneSink}(n,e_i,0)$, (b) ${\sf isOneSink}(n,e_i,1)$, (c) ${\sf isZeroSink}(n,e_i,0)$, and (d) ${\sf isZeroSink}(n,e_i,1)$. 
We assume that $e_i=(v,v')$ and $g$ are subgraphs represented by $n$ with/without $e_i$.
We prove each case if we add $e_i=(v,v')$.
\begin{itemize}
    \item ${\sf isOneSink}(n,e_i,1)$: If the number of terminals connected to $f$, the connectivity satisfies. Here, $g$ satisfies both the tree and minimality conditions; if $g$ has cycles due to $e_i$, new terminals are not connected to $f$ additionally, and $g$ is minimal because $g$ is not Steiner trees without $e_i$.
    \item ${\sf isOneSink}(n,e_i,0)$: ${\sf isOneSink}(n,e_i,0)$ is always false because $g$ does not satisfy the connectivity condition.
    \item ${\sf isZeroSink}(n,e_i,1)$: 
    \begin{itemize} 
    \item If $id_n(v) = id_n(v')$, $g$ has cycle because $v$ and $v'$ are already connected by different paths on the graph. Thus, $g$ does not satisfy the tree condition. 
    \item If $v \notin \frontiers_i$ and $d_n(v)=0$, $v$ becomes a leaf vertex of $g$ after adding $e_i$. If $v$ is not terminals, $g$ could be Steiner tree without $e_i$. Thus, $g$ does not satisfy the minimality condition.  
    \item If $c_n + c(e_i) > \theta$, $g$ does not satisfy cost-constraint.
    \end{itemize}
    \item  ${\sf isZeroSink}(n,e_i,0)$:
    \begin{itemize}
        \item If $v \not\in \frontiers_i \cup \frontiers_{i-1}$, $v$ has a single edge because $v$ does not become frontiers after and before processing $e_i$. Thus, if $v$ is terminals and $e_i$ is deleted, $g$ does not satisfy the connectivity condition.
        \item If $t_n(v)>0$ and $u_n(v) = 1$, $v$ is not connected to other vertices ever. Thus, $g$ does not satisfy the connectivity condition.
        \item If $v \notin \frontiers_i$ and $d_n(v)=1$, $v$ becomes a leaf vertex of $g$ after deleting $e_i$. If $v$ is not terminals, $g$ could be Steiner tree without $e_i$. Thus, $g$ does not satisfy the minimality condition. 
    \end{itemize}
\end{itemize}
\hfill{} $\square$

{\bf Pseudo code for BDD construction method}: 
Algorithm~\ref{alg:construction} shows pseudo code for the BDD consruction method presented in Section 4.2.
The algorithm constructs the BDD by generating nodes from $\nodes_1$ to $\nodes_{|\edges|}$ (line 4).
For each node in $\nodes_{i-1}$, it generates two nodes (lines 5--6).
If nodes satisfy {\sf isOneSink} and {\sf isZeroSink}, node $\alpha$ points at 1- and 0-sinks, respectively (lines 7--10).
Otherwise, it generates $\alpha_x$ by {\sf generateNode} (line 12).
If $\alpha_x'$ is included in $\nodes_i$, it merges $\alpha_x$ and $\alpha_x'$ by {\sf mergeNode} (lines 13-14).
We here note that we use hash functions for efficiently checking whether there is $\alpha_x'$ that satisfy lemma 1.
Finally, we add nodes and arcs to $\nodes_i$ and $\arcs$, respectively (lines 16--17).

\begin{algorithm}[!t] 
	\caption{Construction Algorithm}	\label{alg:construction}
		\DontPrintSemicolon
			    \SetKwInOut{Input}{Input}
	            \SetKwInOut{Output}{Output}
	            \Input{Graph $\graph$, threshold $\threshold$}
	            \Output{BDD $\bdd =(\nodes, \arcs)$}
            	{\bf procedure} {\sc Construction Algorithm}\\
            	order edges;\\
                $\nodes_0 \leftarrow$ $\{ root \}$, $\nodes_i \leftarrow$ $\phi$ for $i=1,2,\ldots, |\edges|$;\\
                 \For{$i=1,2,\ldots, |\edges|$}{
                     \ForEach{$\alpha \in \nodes_{i-1}$}{
                         \ForEach{$x \in \{0,1\}$}{
                            \If{${\sf isOneSink}(n,e_i,x)$}{
                                $\alpha_x \leftarrow$ 1-sink;
                            }
                            \ElseIf{${\sf isZeroSink}(n,e_i,x)$}{
                                $\alpha_x \leftarrow$ 0-sink;
                            }
                             \Else{
                                 $\alpha_x \leftarrow {\sf generateNode}(\alpha,e_i,x)$;\\
                                 \If{$\alpha_x$ and $\alpha_x' \in \nodes_i$ satisfy lemma 1}{
                                     {\sf mergeNode($\alpha_x,\alpha_x'$)};\\
                                }
                                \Else{
                                    $\nodes_i \leftarrow \nodes_i \cup \alpha_x$;\\
                                }
                             }
                             $\arcs \leftarrow \arcs \cup (\alpha, \alpha_x, x)$;\\
                         }
                     }
                 }
              {\bf return} $\bdd$;\\
              {\bf end procedure}
\end{algorithm}

\subsection{BDD Traversal Method}
After constructing BDDs, we traverse the BDD to enumerate Steiner trees whose costs are not larger than a given threshold. 
We first reduce the constructed BDD and then traverse the BDD to obtain the set of Steiner trees.

{\bf BDD reduction}: 
Any decedents of nodes at $i$-th step may not go to 1-sinks even for any patterns of $E^{>i}$.
We can delete these nodes because they are irrelevant to enumerating Steiner trees.
To delete such nodes, we traverse BDD from the 0-sink to the root node.
If both 0-arc and 1-arc of nodes point to the 0-sink, we delete the nodes. Similarly, if both arcs point to the 0-sink or deleted nodes, we also delete the nodes.
This process reduces the size of BDD without sacrificing the accuracy and accelerates the enumeration of Steiner trees.

{\bf Enumerating Steiner trees}: 
We enumerate Steiner trees via traversing the constructed BDDs.
When we traverse the BDD, our algorithm needs to keep costs from the root node to nodes in order to eliminate the Steiner trees with costs larger than $\threshold$.
We traverse the BDD in a breath first search (BFS) manner. In the BFS, we can remove costs at $\nodes_i$ after computing costs at $\nodes_{i+1}$.
Thus, we can reduce the memory costs to keep the costs on nodes.

\section{Scalable extension}
Since the minimal Steiner tree enumeration problem is $\#P$-hard, approximate algorithms are practical in real-world applications as well as the Steiner tree problem.

We need to improve the scalability to large-scale graphs for applying the minimal Steiner tree enumeration problem to the real-world applications.
We use two ideas; (1) restricting the candidates of vertices and edges that can be used for enumerating Steiner trees and (2) enumerating the Top-$k$ Steiner trees (i.e., Steiner trees with $k$ smallest costs) among the set of generated Steiner trees represented by the constructed BDDs.
These two ideas support to enumerate the cost-constrained minimal Steiner trees on large-scale graphs.

\subsection{Preprocessing}
We preprocess graphs before constructing BDD to reduce the size of graphs based on two approaches; {\it seed trees selection} and {\it graph simplification}.
These two techniques effectively reduce the computation cost for constructing BDDs.

{\bf Seed tree selection}: The seed trees are Steiner trees computed by existing algorithms of the Steiner tree problem.
We construct the BDD by the union of seed trees instead of the original graph. 
The benefits of using seed trees are (1) mitigating the computation time and memory usage to construct BDDs and (2) controlling generated Steiner trees by changing the algorithm for selecting seed trees.

{\bf Graph simplification}: We reduce the size of graphs by simplifying the graphs without any information loss. 
We simplify graphs based on the idea of series-parallel graphs \cite{borie2009solving}.
We delete and add the following edges and vertices without sacrificing the number of enumerated Steiner trees as follows:
For sequential edges ($e = (v, v'), e' =(v, v'')$), we delete $v$, $e$ and $e'$, and add a new edge with weights $c(e) + c(e')$ between $v'$ and $v''$, provided that $v$ is not a terminal and its degree is two.
For the loop, we delete the loop because loops are not included in Steiner trees.
We iteratively repeat this process until the graph does not change.

\subsection{Enumerating Top-k Steiner trees}
In the traversal algorithm, we keep the costs from the root to nodes.
The number of paths from the root is significantly large even when BDDs are not very large. 
Thus, we limit the number of costs kept on each node to $k$.
Our algorithm keeps the $k$ smallest costs on each node during traversal, and thus we can guarantee that we obtain the top-$k$ Steiner trees among Steiner trees represented by the constructed BDD.
Our algorithm can control memory usage and computation cost by adjusting $k$.

We describe the time and space complexity of the BDD traversal algorithm.
\begin{theorem}
The time complexity of the BDD traversal algorithm is $O(k|\nodes|\log k)$.
\end{theorem}
{\it Proof}: The BDD traversal algorithm conducts BFS from the root node. The time complexity of BFS is $O(|\nodes|)$ because each node has just two arcs.
To obtain the Steiner trees with $k$ smallest costs, the algorithm sorts costs for each node, which takes $O(k\log k)$.
Therefore, the time complexity of the BDD traversal algorithm is $O(k|\nodes|\log k)$. \hfill{} $\square$

\begin{theorem}
The space complexity of the BDD traversal algorithm is $O(k|\nodes_m|)$, where $|\nodes_m|$ denotes the maximum size of $\nodes_i$ for $i=1,\ldots, |\edges|$.
\end{theorem}
{\it Proof}: The BDD traversal algorithm keeps $k$ costs for each node. 
Since the algorithm conducts BFS, it removes $k$ costs at $\nodes_i$ after computing costs at $\nodes_{i+1}$.
It stores at most $k \cdot |\nodes_m|$ costs during the traversal.
Thus, the space complexity of the BDD traversal algorithm is $O(k|\nodes_m|)$.
\hfill{} $\square$

We here note that our algorithm enumerates top-k $k$ Steiner trees and in addition, it possibly enumerates other Steiner trees whose costs are not larger than the threshold.








\section{Experimental Study}\label{sec:experiment}

We conduct experimental studies to evaluate our scalable method.
Our goal in the experiments is to answer the following questions:
{\it Efficiency}. Does our algorithm efficiently enumerate Steiner trees?
{\it Accuracy}. Does our algorithm find Steiner trees with small costs? and
{\it Scalability}. Does our algorithm handle large-scale graphs?
We here note that our algorithm without scalable extensions did not work for large-scale graphs.

Our algorithms were implemented in C++.
Experiments were performed on a Linux server with 512GB of memory and an Intel(R) Xeon(R) CPU E5-2699v3 @ 2.30GHz processor. All algorithms are single-threaded.

\subsection{Experimental Setting}
We provide an overview of our experimental setup, including
datasets, comparisons, and parameters.

\begin{table}[ttt]
    \centering
    \caption{Datasets}
    \label{tab:datasets}
    \begin{tabular}{|c|rrrr|}\hline
         Name &\multicolumn{1}{c}{$|\vertices|$}&\multicolumn{1}{c}{$|\edges|$}&\multicolumn{1}{c}{$|\terminals|$}&\multicolumn{1}{c|}{Min cost}\\ \Hline         
         {\bf PUCNcc3-5n}  & 125& 750& 13 & 20 \\
         {\bf PUCNcc5-3n}  & 243 & 1{,}215 & 27  & 42 \\
         {\bf PUCNcc6-3n}  & 729 & 4{,}368& 76 & 100 \\
         {\bf ALUE2087}  & 1{,}244& 1{,}971& 34 & 1{,}049 \\
         {\bf ALUE2105}  & 1{,}220& 1{,}858& 34 & 1{,}032 \\
         {\bf ALUE3146}  & 3{,}636& 5{,}869& 64 & 2{,}240 \\
         {\bf G101}  & 67{,}966& 82{,}485& 100 & 3{,}492{,}405 \\
         {\bf G105}  & 79{,}244& 101{,}189& 550 & 12{,}507{,}877 \\
         {\bf G301}  & 80{,}736& 98{,}750& 191 & 4{,}797{,}441 \\
         \hline
    \end{tabular}
\end{table}

\begin{table*}[ttt]
    \centering
    \caption{Efficiency and accuracy. DNF indicates that the tests did not finish over 24 hours.}
    \label{tab:overviewresult}
    {\small
    \begin{tabular}{|c|rrr|rrr|rrr|}\hline
              &\multicolumn{3}{c|}{Our algorithm}&\multicolumn{3}{c|}{ToSP}&\multicolumn{3}{c|}{Scip-jack}\\
         Data &\multicolumn{1}{c}{Time} &Avg cost&Min cost&Time&Avg cost&Min cost&Time &Avg cost&Min cost\\ \Hline
         {\bf PUCNcc3-5n}  &1.23 s  &23.6 &20&290.7 s &22.0 &20  & DNF &--- &---\\
         {\bf PUCNcc5-3n}  &7.49 s  &50.9 & 48&80.1 s &49.2 &47  & DNF &--- &---\\
         {\bf PUCNcc6-3n}  &1{,}623.4 s  &101.0 &101 & 8315.1 s& 119.8 &116 & 1{,}218.3 & 108.4 & 102\\
         {\bf ALUE2087}   &0.07 ms &1{,}070.7 &1{,}049& 0.02 s& 1{,}142.8 &1{,}104 &0.85 s&1{,}059.3 &1{,}049\\
         {\bf ALUE2105}   &0.16 ms  &1{,}080.9&1{,}032&0.25 s&1{,}129.0 &1{,}097 &18.6 s& 1{,}049.7 & 1{,}032\\
         {\bf ALUE3146}  &34.5 s & 2{,}267.9&2{,}240 &990.1 s& 2{,}599.7&2{,}438 &DNF& --- &---\\
         {\bf G101}  & 0.03 s&   3{,}812{,}251 &3{,}801{,}188& 9.5 s& 3{,}859{,}434 & 3{,}726{,}148&423.1 s&3{,}583{,}357.0 &3{,}515{,}778\\
         {\bf G105}  & 347.6 s &   13{,}149{,}050&13{,}146{,}451 & 2{,}382.1 s&  13{,}358{,}044 & 13{,}134{,}651 &85{,}971.3 s &12{,}731{,}646.7 &12{,}596{,}785\\
         {\bf G301}  & 730.7 s&   5{,}113{,}457&5{,}110{,}959 & 1{,}220.3 s& 5{,}270{,}464 & 5{,}145{,}846 &37{,}619.0 s& 201{,}715.4&4{,}847{,}830\\\hline
    \end{tabular}
    }
\end{table*}

\begin{figure*}[ttt]
\centering
\begin{minipage}[t]{1.0\linewidth}
    \centering
    \includegraphics[width=1.0\linewidth]{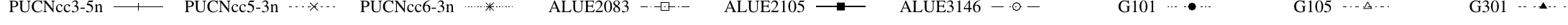}
    \end{minipage}
    \\\vspace{-9mm}
\begin{minipage}[t]{1.0\linewidth}
    \centering
  \subfloat[Graph size]{\epsfig{file=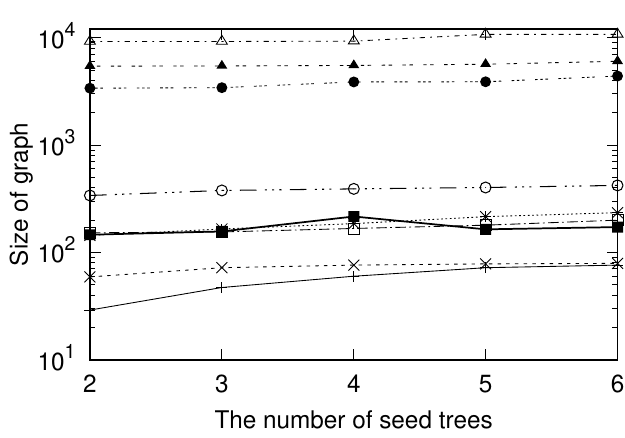,width=0.32\linewidth}}
  \subfloat[BDD size]{\epsfig{file=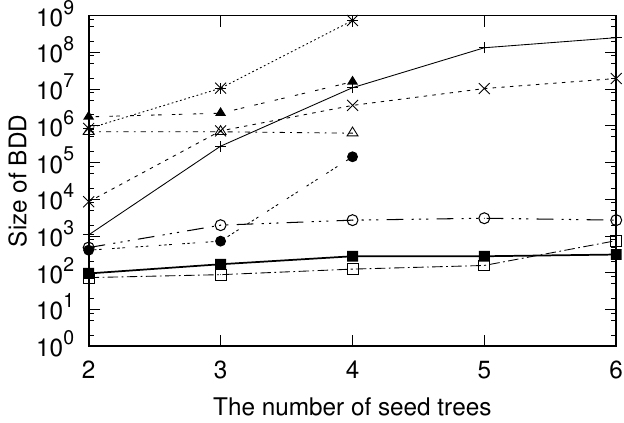,width=0.32\linewidth}}
  \subfloat[Construction Time]{\epsfig{file=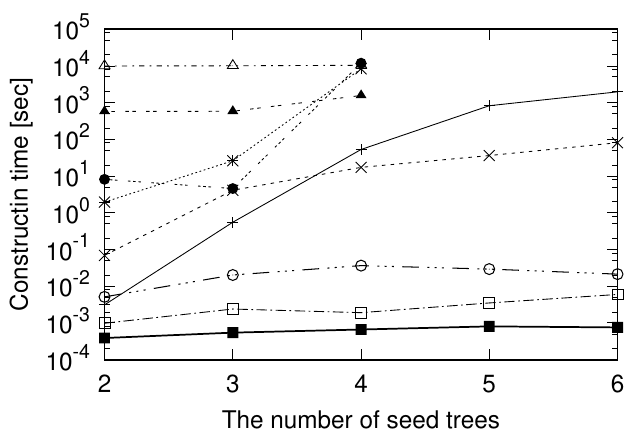,width=0.32\linewidth}}
  \caption{Scalability: varying the number of seed trees. Missing plots are due to out of memory.}
  \label{fg:scalability}
	\end{minipage}
\end{figure*}

{\bf Datasets}: 
We use three types of datasets; code covering\footnote{http://steinlib.zib.de/showset.php?PUC}, VLSI\footnote{http://steinlib.zib.de/showset.php?ALUT}, telecommunication networks\footnote{https://homepage.univie.ac.at/ivana.ljubic/research/STP/}.
The datasets include graphs, terminals, and the minimum cost of the Steiner tree for the given graphs and terminals.
We use PUCNcc3-5n, PUCNcc5-3n, PUCN6-3n from code covering, ALUE2087, ALUE2105, and ALUE3146 from VLSI dataset, and G101, G105, and G301 from telecommunication networks.
Table \ref{tab:datasets} shows statistics of the graphs.

{\bf Comparisons}: 
We compare our scalable method with algorithms by using the tree-of-shortest-path (ToSP for short) approximation \cite{thomas2001introduction,keyder2009trees} and SCIP-Jack solver \cite{gamrath2017scip}\footnote{https://github.com/dRehfeldt/scipjack}, respectively.
The ToSP approximation computes a Steiner tree by computing the shortest paths from a terminal to other all terminals, which is efficient without guaranteeing accuracy of approximation.
SCIP-Jack can compute Steiner trees accurately more than the ToSP algorithm, but it takes a more cost.
To enumerate Steiner trees, we repeat (1) temporally deleting $x \%$ edges randomly from the given original graph, (2) running the algorithms on the graph, and (3) checking whether duplicate trees are already generated, until obtaining the specified number of Steiner trees with costs not larger than the given threshold.
We specify the number of Steiner trees to the number of Steiner trees that our algorithm obtains.
We here note that we can use other approximate algorithms and solvers. 

{\bf Parameters}: 
We set parameters to our algorithms; threshold $\threshold$, the number of seed trees, and the rank $k$.
Threshold $\theta$ is 1.2 times the cost of optimal Steiner trees. 
We set the number of seed trees is three as default, and we compute the seed trees by using the comparisons.
We decide $k$ from [1{,}000, 10{,}000, 100{,}000, and 1{,}000{,}000] to enumerate Steiner trees as much as possible without out of memory, based on our preliminary experiments.
To order edges for constructing BDDs, we use a BFS from the terminal with the smallest degree.
For the comparisons, we decide $x$ from [1, 5, 10, 20] in each test, and we report the results with the smallest computation time.


\subsection{Result}
We first show the efficiency and accuracy of our algorithm and comparisons.
Table \ref{tab:overviewresult} shows the enumeration time, the average cost of enumerated Steiner trees, and the minimum cost among enumerated Steiner trees on each method.


Our algorithm finds Steiner trees more efficiently than the ToSP and Scip-Jack, except for PUCNcc6-3n.
Since our BDDs compactly represent the set of Steiner trees, we can find Steiner trees without duplication checks.
The comparisons are inefficient because they often find duplicate Steiner trees and generate trees whose costs larger than the threshold. 
The enumeration time on our algorithm is related to the size of BDD. 
If the BDD is small, the traverse of BDDs is quickly terminated.
Otherwise, it takes a large time such as PUCNcc6-3n.

We can see that the accuracy of our algorithm is better than that of ToSP in most datasets.
This is because our algorithm guarantees enumerating the Steiner trees with the $k$ smallest costs among them represented by the BDD.
The Scip-Jack is more accurately than our algorithm except for PUCNcc6-3n.
Scip-Jack finds sub optimal Steiner trees for the given graph, but it takes a large time to enumerate Steiner trees.
In PUCNcc6-3n, the BDD represents numerous Steiner trees, so it can enumerate Steiner trees with very small costs.


Finally, we show the scalability of our algorithm.
The scalability of our algorithm is related to the number of seed trees. Therefore, we evaluate the impacts on the number of seed trees.
Figure \ref{fg:scalability} shows the size of graphs, the size of BDD, and construction time, with varying the number of seed trees.

The size of graphs slightly increases as the number of seed trees increases. The size does not increase much because seed trees have many overlapped edges. Though, the BDD size and construction time significantly increase. 
From these results, we can know that it is necessary to control computation costs and memory usage.
Since our algorithm can control them by changing the number of seed trees, it is applicable to large-scale graphs.

From these results, we confirm that our algorithm efficiently enumerates minimal Steiner trees whose costs are close to the optimal costs, with high scalability.

\section{Related Work}\label{sec:related}

We here describe existing works related to the Steiner tree enumeration problem and binary decision diagram.

{\bf Steiner Tree Enumeration Problem}: 
There are several known results on algorithms related to Steiner tree enumeration and its variants~\cite{kimelfeld2006finding,khachiyan2005complexity,dourado2014algorithmic,conte2019listing,kimelfeld2008efficiently}.

Dourado et al.\  \cite{dourado2014algorithmic} and Khachyan et al.\ \cite{khachiyan2005complexity} 
theoretically studied {\it minimum} Steiner tree enumeration problem and Steiner forest, respectively.
In \cite{kobayashi2020polynomial,kimelfeld2006finding,kimelfeld2008efficiently}, they proposed algorithms for enumerating minimal Steiner trees and analyzed computational complexity of their algorithms. 
Their algorithms interatively remove/add edges from/to the given graph to enumerate minimal Steiner trees.
Unfortunately, the algorithms obviously do not work if graphs are large scale and the number of terminals is large, so they are not practical. 
For example, Kimelfeld and Sagiv \cite{kimelfeld2006finding} proposed an algorithm for enumerating minimal Steiner trees with rankded order, which takes $O(4^{|T|}|V|+3^{|T|}e \log |V|)$ for generating a single Steiner tree, where $|T|$ and $|V|$ are the numbers of terminals and vertices, respectively.
All existing works have not shown that their algorithms are empirically applicable to real-world graphs.

{\bf Subgraph Enumeration by Binary Decision Diagram}: 
Binary decision diagram (BDD) \cite{bergman2016decision} as well as its variants such as ZDD \cite{minato1993zero} and SDD \cite{choi2013dynamic} are effective to enumerate discrete data structures.

One of the main research challenges in the BDD is efficient construction of BDDs tailored for specific problems.
For example, Bergman et al.\ \cite{bergman2011manipulating} and Maehara et al.\ \cite{maehara2017exact} proposed construction algorithms for the set covering and influence spread problems, respectively.
The BDD is often used to enumerate constrained subgraphs, such as paths, spanning trees, and connected graphs \cite{yan2015novel,hardy2005computing,sasaki2019efficient,maehara2017exact}.
However, to the best of our knowledge, BDDs for enumerating minimal Steiner trees are not studied, and existing methods are not easily applied to the minimal Steiner tree enumeration problem. In addition, these methods are not scalable to the size of graphs.
Thus, our method is the first scalable method for enumerating the minimal Steiner trees.

Enumerating the cost-constrained subgraphs could have many applications.
However, to the best of our knowledge, there are no BDD construction methods for enumerating the cost-constrained subgraphs. 
Our approach can be used for other problem definitions.

Our approach with scalable extension is different from existing works (e.g., \cite{bergman2011manipulating}).
We aim to constructs BDDs not including subgraphs that are not Steiner trees.
Our approach resizes the given graphs and enumerates the top-k Steiner trees. 
Existing approximate BDD construction methods construct BDDs that possibly include subgraphs that are not targeted ones because they approximately merge/delete nodes of BDDs.
Therefore, our approach is novel compared with other BDD methods.

\section{Conclusion} \label{sec:conclusion} 
We initiated a new problem, the cost-constrained minimal Steiner tree enumeration problem.
For addressing the problem, we proposed a novel algorithm based on the binary decision diagram.
We showed that our algorithm is efficient, accurate, and scalable through experimental studies using real-world datasets.


\bibliographystyle{named}
\bibliography{bddsteiner}

\end{document}